\documentclass[12pt]{article}
 \usepackage[margin=.9in,footskip=0.25in]{geometry}
 \usepackage{amssymb}
\usepackage[T1]{fontenc}
\usepackage[latin9]{inputenc}
\usepackage{array}
\usepackage{booktabs}
\usepackage{mathrsfs}
\usepackage{amsmath}
\usepackage{graphicx}
\usepackage{esint}
\usepackage{bm}
\usepackage[symbol]{footmisc}
\usepackage{textcomp}
\usepackage{wrapfig}
\usepackage[sort&compress, numbers]{natbib}
\usepackage{hyperref}
\usepackage{mathtools}
\usepackage{authblk}
\usepackage{setspace}
\begin{document}

\title{Compactly supported travelling waves arising in a general reaction-diffusion Stefan model}

\footnotesize
\author{Nabil~T.~Fadai$^*$}
\affil{\footnotesize{School of Mathematical Sciences, University of Nottingham, Nottingham NG7 2RD, United Kingdom.
\\
*Corresponding author email address: \texttt{nabil.fadai@nottingham.ac.uk}
}} 

\maketitle
\normalsize

\begin{abstract}
We examine travelling wave solutions of the reaction-diffusion equation, $\partial_t u= R(u) +  \partial_x \left[D(u) \partial_x u\right]$, with a Stefan-like condition at the edge of the moving front. With only a few assumptions on $R(u)$ and $D(u)$, a variety of new compactly supported travelling waves arise in this Reaction-Diffusion Stefan model. While other reaction-diffusion models admit compactly supported travelling waves for a unique wavespeed, we show that compactly supported travelling waves in the Reaction-Diffusion Stefan model exist over a range of wavespeeds. Furthermore, we determine the necessary conditions on $R(u)$ and $D(u)$ for which compactly supported travelling waves exist for all wavespeeds. Using asymptotic analysis in various distinguished limits of the wavespeed, we obtain approximate solutions of these travelling waves, agreeing with numerical simulations with high accuracy.

\end{abstract}

\section{Introduction}

Travelling waves arise in a wide range of reaction-diffusion models \cite{murray, sherratt1990models, aronson1980density, witelski1995merging}, including in ecology \cite{fisher1937wave, kolmogorov1937etude, maini2004traveling, du2010spreading, bao2018free}, cell biology \cite{el2019revisiting, fadai2020new, mccue2019hole, krause2019influence}, and industrial applications involving heat and mass transfer \cite{mcguinness2000modelling, dalwadi2017mathematical, fadai2018asymptotic, brosa2019extended}. These reaction-diffusion models are useful for describing the concentration of a particular species, $u(x,t)$, in which the travelling wave moves at a constant wavespeed, $c$, over the one-dimensional domain $x\in\mathbb{R}$ at time $t>0$.
Notably, there has been particular interest in travelling waves that exhibit \textit{compact support}, in which $u(x,t)$ is a monotone decreasing function that terminates at $u=0$ when at some finite spatial position $x(t)$. These travelling waves are useful in applications where a well-defined ``edge'' of a travelling wave is an essential component of the modelling framework \cite{mccue2019hole, el2019revisiting, fadai2020new}. While many existing reaction-diffusion models have been shown to produce a compactly-supported travelling wave for a single wavespeed, $c^*$ \cite{sherratt1990models, sherratt1996nonsharp, aronson1980density, witelski1995merging, garduno1994approximation, sanchezgarduno1995traveling, de1991travelling, de1998travelling, needham1999reaction}, there are only a handful of modelling frameworks giving rise to compactly-supported travelling waves for \textit{range} of wavespeeds.

One particular reaction-diffusion modelling framework giving rise to compactly-supported travelling waves over a range of wavespeeds is via the incorporation of a {moving boundary} \cite{du2010spreading, bao2018free, el2019revisiting, fadai2020new},  whereby $x\in(-\infty,L(t)]$ and $L(t)$ evolves based on a Stefan-like condition at the edge of travelling wave. For particular choices of linear \cite{du2010spreading, bao2018free, el2019revisiting} and degenerate diffusivities \cite{fadai2020new}, $D(u)$, travelling waves with compact support exist for \textit{all} wavespeeds $c\in[0,c^*]$, where the value of the critical wavespeed $c^*$ depends on $D(u)$. However, generalisations of the models presented in \cite{el2019revisiting, fadai2020new} for a broader class of reaction functions, $R(u)$ and nonlinear diffusivities, $D(u)$, has yet to be considered.

In this work, we consider a general reaction-diffusion model with a moving boundary, which we refer to as the \textit{Reaction-Diffusion-Stefan} (RDS) model. Specifically, we determine the necessary conditions for $R(u)$ and $D(u)$ so that travelling waves exist with compact support over a range of wavespeeds.  Using asymptotic analysis in the limit where $c\ll1$, we obtain approximate solutions of these travelling waves, agreeing with numerical simulations with high accuracy. Additionally, we determine the approximate relationship between $c$ and the Stefan parameter, $\kappa$, which relates the concentration flux, $-D(u)\partial_{x}u(x,t)$, with the speed of the moving boundary. Along with the asymptotic analysis performed when $c\ll1$, corresponding to when $\kappa\ll 1$, we also consider approximate solutions of travelling waves for $\kappa\gg1$. Depending on the kinetics of $R(u)D(u)$ near $u=0$, we show that having $\kappa\gg 1$ is equivalent to $c$ approaching a finite wavespeed, or that $c\gg1$. In particular, we outline the necessary conditions for $R(u)$ and $D(u)$ so that travelling waves exist with compact support for \textit{all} wavespeeds (i.e., $c^*=\infty$). For both $c^*$ finite and infinite, we determine asymptotic approximations for both the travelling wave and the $\kappa(c)$ relationship in the limit where $\kappa\gg1$.

\section{Travelling waves in the Reaction-Diffusion-Stefan model}
We consider the following non-dimensional reaction-diffusion model, describing the concentration $u\in[0,1]$, on the spatial domain $x\in(-\infty, L(t)]$ with a Stefan-like condition at the moving boundary $x=L(t)$:
\begin{align}
&\partial_t u(x,t)= R(u) +  \partial_{x}\left[D(u)\partial_{x}u(x,t)\right], \label{eq:PFS}
\\
&\lim_{x\to-\infty} u(x,t)=1,
&u(L(t),t)=0,
\\
&\frac{\mathrm{d}L}{\mathrm{d}t} = -\kappa \left. D(u)\partial_{x}u(x,t)\right|_{x\to L(t)^-}, 
&L(0)=L_0.\label{eq:PFSa}
\end{align}
The Stefan condition relates the speed of the moving front, $\displaystyle  \mathrm{d}L/\mathrm{d}t$, to the concentration flux, $-D(u)\partial_{x}u(x,t)$ via the Stefan parameter, $\kappa$. We will also impose that 
\begin{equation}
\begin{aligned}
0 < D(u) <\infty, \qquad 0 &\le R(u) <\infty, \quad u\in (0,1], 
\\
 R(u)D(u) &= o(u^{-1}), \quad u \to 0^+. 
\end{aligned}\label{eq:bounds}
\end{equation}
These bounds indicate that degenerate diffusion, i.e. $D=0$ or $D=\infty$, can only occur if $u=0$, and that the reaction  $R(u)$ can only be a source term.

The main solutions of the system \eqref{eq:PFS}--\eqref{eq:PFSa} that we will examine are travelling wave solutions. Therefore, we transform the PDE system into travelling wave coordinates via $z = x-L_0-ct,$ where $z\in (-\infty,0]$ and $c\ge0$. Noting that when $x=L(t)$, we have that $L(t)=L_0+ct$, and hence,
$\mathrm{d}L/\mathrm{d}t = c$. By denoting the concentration flux as
\begin{equation}
Q(z) = D(u)\frac{\mathrm{d}u}{\mathrm{d}z}, \label{eq:flux}
\end{equation}
we obtain, by transforming \eqref{eq:PFS} into travelling wave coordinates and multiplying by $D(u)$,
\begin{equation}
D(u)\frac{\mathrm{d}Q}{\mathrm{d}z} =-cQ(z)-R(u)D(u).\label{eq:flux2}
\end{equation}
Provided that $D(u)\mathrm{d}u/\mathrm{d}z \to 0$ as $u\to 1^-$, the system of first-order nonlinear differential equations \eqref{eq:flux}--\eqref{eq:flux2} is coupled to a set of boundary conditions arising from \eqref{eq:PFSa}:
\begin{equation}
\lim_{z\to-\infty} u(z) = 1, \qquad \lim_{z\to-\infty} Q(z) = 0, \qquad  u(0) = 0, \qquad  Q(0)=-\frac{c}{\kappa}. \label{eq:BC}
\end{equation}
We will focus on the heteroclinic trajectory $(u,Q)$ phase plane beginning at $(1,0)$ and terminating at $(0,-c/\kappa)$ to agree with the boundary conditions listed above. In doing so, we determine not only the relationship between $Q$ and $u$, but also the relationship between $c$ and $\kappa$. Therefore, by dividing \eqref{eq:flux2} by \eqref{eq:flux}, we have that
\begin{equation}
\begin{aligned}
&-\frac{\mathrm{d}Q(u)}{\mathrm{d}u} = c+\frac{R(u)D(u)}{Q(u)}, 
\\
&Q(1)=0, \qquad \lim_{u\to0^+} Q(u)=-\frac{c}{\kappa}. 
\end{aligned}\label{eq:dvdu}
\end{equation}

\subsection{Some explicit solutions of the heteroclinic trajectory $Q(u)$}
While by no means an exhaustive list, there are some straightforward solutions of \eqref{eq:dvdu} for particular choices of $R(u)D(u)$. In doing so, we are also  able to obtain the explicit relationship between $c$ and $\kappa$. The simplest case to consider is when $R(u)D(u) = A$, where $A>0$. In this case, \eqref{eq:dvdu} becomes a separable differential equation whose solution is
\begin{equation}
R(u)D(u)\equiv A \implies Q(u) = -\frac{A}{c}\left\lbrace1+W\left[-
\exp\left(-1-\frac{c^2(1-u)}{A}\right)\right]\right\rbrace, \label{eq:cst}
\end{equation} 
where $W(\cdot)$ is the Lambert-W function \cite{corless1996lambertw}. Therefore, by evaluating this expression at $u=0$, we retrieve the relationship between $c$ and $\kappa$:
\begin{equation}
R(u)D(u)\equiv A \implies \kappa = \frac{c^2}{A\left\lbrace1+W\left[-
\exp\left(-1-\frac{c^2}{A}\right)\right]\right\rbrace}.
\end{equation}
However, since $A>0$, we have, from \eqref{eq:dvdu} that $\mathrm{d}Q(u)/\mathrm{d}u \to -\infty$ as $u\to 1^-$. Furthermore, since $D(1)$ is positive and finite, the travelling wave $u(z)$ will evolve on some compact interval $[Z_0,0]$ and $u(z)\equiv 1$ for $z\le Z_0$. 
Should a travelling wave be required to evolve on the semi-infinite interval $(-\infty,0]$, we must further impose that $R(1)=0$. However, as we will show in Section 2.4, having $R(1)=0$ is a necessary but not sufficient condition in obtaining travelling waves on the semi-infinite interval $(-\infty,0]$.

A simple reaction-diffusion model that exhibits travelling waves on a semi-infinite domain is
 $R(u)D(u)=A(1-u)$, where $A>0$. Solving \eqref{eq:dvdu} for this choice of $R(u)D(u)$ yields
\begin{equation}
R(u)D(u)\equiv A(1-u) \implies Q(u) =\frac{ \left(\sqrt{c^2+4A}-c\right)(u-1)}{2}
, \qquad \kappa = \frac{c\left(\sqrt{c^2+4A}+c\right)}{2A}.
\end{equation}
Therefore, as $u\to1^-$, $\mathrm{d}u/\mathrm{d}z \propto u-1$, confirming that travelling waves in this reaction-diffusion model decay exponentially to $u=1$ as $z\to-\infty$.

\section{Approximating Reaction-Diffusion-Stefan travelling waves with small wavespeed}
\subsection{Approximating the heteroclinic trajectory $Q(u)$ for $c\ll 1$}
For more general $R(u)D(u)$ function forms, an explicit solution of  \eqref{eq:dvdu} is not always possible. Instead, we consider the solution of \eqref{eq:dvdu} first in the limit where $0<c\ll 1$ by expanding $Q(u)$ as a regular perturbation expansion in $c$, i.e. 
$Q(u) = q_0(u) + c q_1(u) + \mathcal{O}(c^2)$.
Substituting this expansion into \eqref{eq:dvdu} provides the following two-term equations:
\begin{align}
&\mathcal{O}_s(1): \quad -q_0(u) \frac{\mathrm{d}q_0(u)}{\mathrm{d}u} = R(u)D(u),   \quad &q_0(1)=0; \label{eq:O1}
\\
&\mathcal{O}_s(c): \quad -\frac{\mathrm{d}}{\mathrm{d}u}[q_0(u) q_1(u)] =q_0(u), \quad &q_1(1)=0. \label{eq:Oc}
\end{align}
Solving \eqref{eq:O1}--\eqref{eq:Oc} yields
\begin{equation}
Q(u) =D(u)\frac{\mathrm{d}u}{\mathrm{d}z}\sim -\xi(u) + \frac{c}{\xi(u)}\int_u^1 \xi(v)\mathrm{d}v, ~~\text{where}~~ \xi(u) = \sqrt{\int_u^1 2R(s)D(s)\mathrm{d}s}. \label{eq:twoterm}
\end{equation}
From \eqref{eq:bounds}, we have that $\xi(u)$ is non-negative, bounded, and monotone decreasing for $u\in[0,1]$.
Furthermore, evaluating \eqref{eq:twoterm} at $u=0$ provides a two-term approximation for the wave speed $c$ as a function of the Stefan parameter $\kappa$ when $c\ll 1$:
\begin{equation}
\displaystyle\kappa \sim  \frac{c\, \xi(0)}{ \xi(0)^2-c\int_0^1 \xi(s)\mathrm{d}s}. \label{eq:ck0}
\end{equation}
Therefore, we can obtain two approximate relationships in the limit where the wavespeed is small: the relationship between the travelling wave and its flux, as well as the relationship between the wavespeed and the Stefan parameter.

\subsection{Approximating the travelling wave $u(z)$ for $c\ll 1$ and $u\to0^+$}

From \eqref{eq:twoterm}, we have an approximate solution for the heteroclinic trajectory describing the travelling wave in the $(u,Q)$ phase plane. It remains to show how to construct an approximate solution for $u(z)$ from this trajectory. We first examine the behaviour of the travelling wave near its leading edge, i.e. for $u\to 0^+$.
By approximating $R(u)D(u) \sim B u^q$ for $u\to0^+$, where $q>-1$, we have that
\begin{equation}
\xi(u) \sim \xi(0) - \frac{Bu^{q+1}}{\xi(0)(q+1)}
~~\text{ and }~~
\int_u^1 \xi(v)\mathrm{d}v \sim \int_0^1 \xi(v)\mathrm{d}v  - \xi(0)u. \label{eq:xiapx}
\end{equation}
By substituting \eqref{eq:xiapx} into \eqref{eq:twoterm}, we obtain
\begin{equation}
-\frac{D(u)}{\xi(0)}\frac{\mathrm{d}u}{\mathrm{d}z}\sim 1- \frac{Bu^{q+1}}{(q+1)\xi(0)^2} -c\omega\left( 1-\frac{u}{\xi(0)\omega}+ \frac{Bu^{q+1}}{(q+1)\xi(0)^2} 
\right),\label{eq:ODEapx}
\end{equation} 
where
\begin{equation}
\omega = \frac{1}{\xi(0)^2} \int_0^1 \xi(v)\mathrm{d}v. 
\end{equation}
Therefore, in the limit where $u\to0^+$ and $c\ll1$, we have, by employing the initial condition $u(0)=0$, that
\begin{equation}
-\xi(0)z \sim \int_0^u D(s)\left[1+ \frac{Bs^{q+1}}{(q+1)\xi(0)^2} +c\omega\left( 1-\frac{s}{\xi(0)\omega}+ \frac{Bs^{q+1}}{(q+1)\xi(0)^2} \right)\right]\mathrm{d}s. \label{eq:intD}
\end{equation}
Crucially, we note that $D(u)$ must be $o(u^{-1})$ for the right-hand side of \eqref{eq:intD} to converge, thereby producing compact support for the travelling wave near its leading edge. By approximating $D(u)\sim Au^{p}$ for $u\to0^+$, where $p>-1$, we have that the leading edge of the travelling wave can be determined implicitly as
\begin{equation}
\begin{aligned}
z\sim -\frac{Au^{p+1}}{(p+1)\xi(0)} &\left[1+ \frac{B(p+1)u^{q+1}}{(q+1)(p+q+2)\xi(0)^2} \right.
\\
&\left. +c\omega\left( 1-\frac{(p+1)u}{(p+2)\xi(0)\omega}+ \frac{B(p+1)u^{q+1}}{(q+1)(p+q+2)\xi(0)^2} \right)\right]. 
\end{aligned}\label{eq:Usmall}
\end{equation}
Thus, by knowing the approximate expansions of $D(u)$ and $R(u)D(u)$ near $u=0$, we are able to approximate $u(z)$ for $u\to0^+$.

\subsection{Approximating the travelling wave $u(z)$ for $c\ll 1$ and $u\to1^-$}
Having approximated the travelling wave $u(z)$ near its leading edge, i.e. when $u\ll 1$, we now examine the far-field behaviour of $u(z)$ by considering the limit where $u\to1^-$. From the conditions imposed in \eqref{eq:bounds}, we have that $D(1)$ is strictly positive and finite, as well as $R(u)D(u) \sim \beta(1-u)^\gamma, \gamma\ge0,$ for $u\to1^-$ . With these approximations, \eqref{eq:twoterm} becomes
\begin{equation}
D(1) \frac{\mathrm{d}u}{\mathrm{d}z}\sim -\sqrt{\frac{2\beta}{\gamma+1}}(1-u)^{\frac{\gamma+1}{2}}
+\frac{2c(1-u)}{\gamma+3}. \label{eq:FF}
\end{equation}
We note that in the special case where $\gamma=1$, $\mathrm{d}u/\mathrm{d}z \propto 1-u$ at leading order, implying that
\begin{equation}
\gamma=1 \implies u\sim 1-\exp\left[\left(\sqrt{\beta}-\frac{c}{2}\right)\left(\frac{z-Z_0}{D(1)}\right)\right], \label{eq:gam1}
\end{equation}
where $Z_0$ is an arbitrary constant. Next, when $0\le \gamma<1$, we rearrange \eqref{eq:FF} and obtain
\begin{equation}
\int \frac{\mathrm{d}u}{ (1-u)^{\frac{1+\gamma}{2}}\left[1-\frac{2c}{\gamma+3}\sqrt{\frac{\gamma+1}{2\beta}}(1-u)^{\frac{1-\gamma}{2}}\right]} \sim -\sqrt{\frac{2\beta}{\gamma+1}} \left(\frac{z-Z_0}{D(1)}\right). \label{eq:intcompact}
\end{equation}
Since $\gamma<1$, the bracketed term in the integrand in \eqref{eq:intcompact} is strictly positive for $u\in[0,1]$ and is close to 1 for $c\ll1$ and $u\to1^-$. Therefore, we obtain the implicit far-field approximation 
\begin{equation}
z \sim Z_0+ \left(\frac{2D(1)}{1-\gamma}\right)\left[\sqrt{\frac{\gamma+1}{2\beta}}(1-u)^{\frac{1-\gamma}{2}}+\frac{c(1-u)^{1-\gamma}}{\gamma+3}\right], \qquad 0\le \gamma<1. \label{eq:gam0}
\end{equation}
Furthermore, we note that when $u=1$, $z=Z_0,$ indicating that when $0\le\gamma<1$, travelling waves will have a \textit{finite} compact support interval. 
Finally, we examine the case where $\gamma>1$.  From \eqref{eq:intcompact}, we identify that the integrand will now have two singularities in $[0,1]$:
\begin{equation}
u=1  ~~\text{and}~~ u=1-\left[\frac{2c}{\gamma+3}\sqrt{\frac{\gamma+1}{2\beta}}\right]^{2/(\gamma-1)}. 
\end{equation}
Therefore, for the travelling wave to deviate significantly away from $u=1$, we must have $c\equiv0$ when $\gamma>1$. This key result indicates that in order for travelling waves in the Reaction-Diffusion-Stefan model to have a semi-infinite domain with non-zero wavespeed, then $R(u)=\mathcal{O}_s(u-1)$ as $u\to1^-$.

\subsection{Comparison between asymptotic approximations and numerical solutions}\label{sec:smallc}
Now knowing the approximate forms of the travelling wave, $u(z)$, near its leading edge and near $u=1$ for small wavespeeds, we compare the numerically-determined travelling waves with these approximations for various reaction-diffusion models. To compute the numerical solution of \eqref{eq:flux}--\eqref{eq:BC} for various $R(u)D(u)$ and $c$ inputs, we use \texttt{ode15s} in MATLAB.

\begin{figure}
\begin{center}
\includegraphics[width= .9\textwidth]{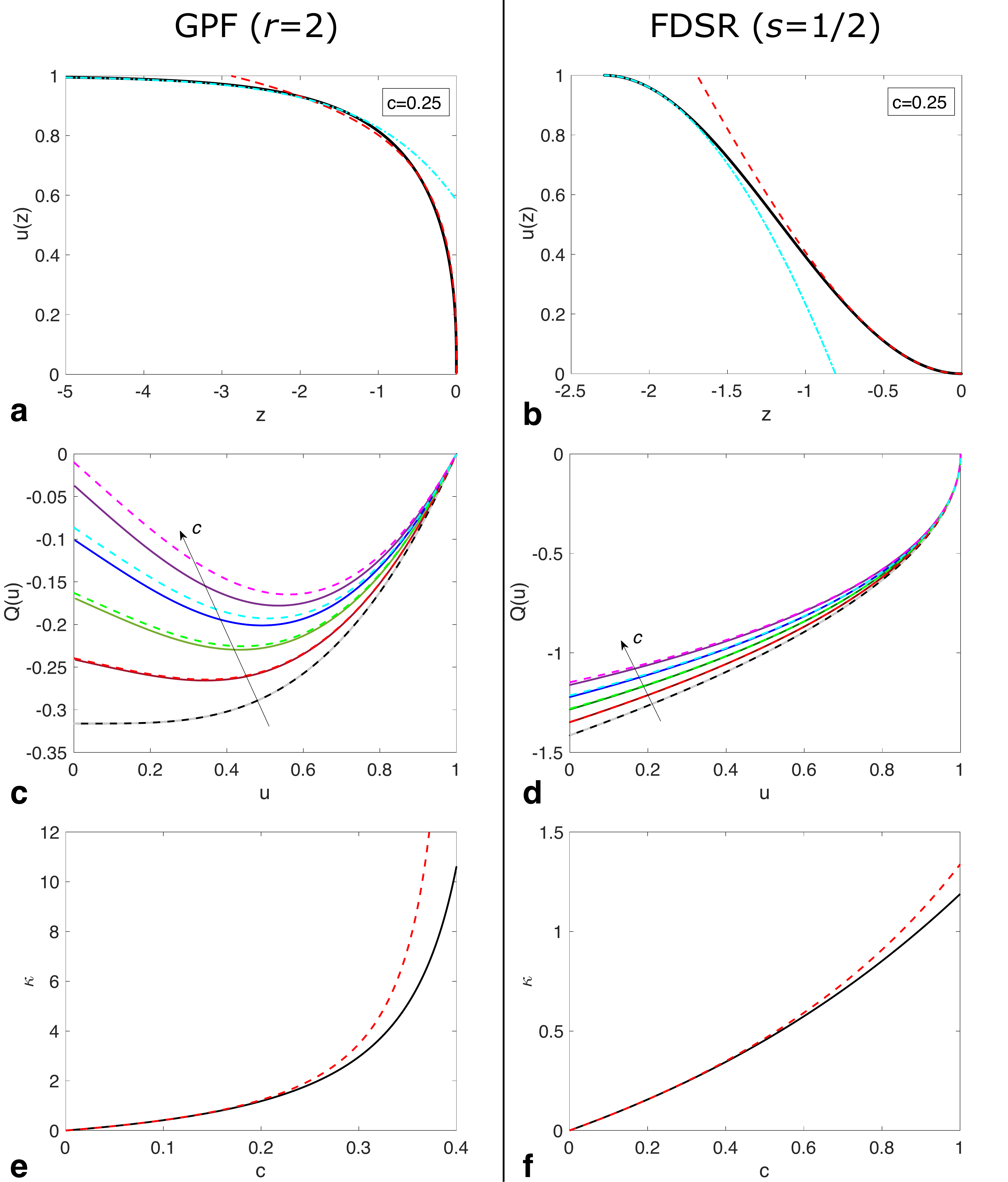}
\end{center}
\vspace{-0.2in} 
 \caption{\textbf{Travelling waves arising in two models of the RDS model and asymptotic approximations valid for $c\ll1$.} Left: GPF model \eqref{eq:GPF} with $r=2$; right: FDSR model \eqref{eq:FDSR} with $s=1/2$. (a,b) Numerically-computed travelling waves (black curve) compared with approximations for $u\to0^+$ \eqref{eq:Usmall} (red dashed curve) and $u\to1^-$ \eqref{eq:GPFz},\eqref{eq:FDSRz} (cyan dot-dashed curve), with $c=0.25$. (c,d) Comparisons of the numerically-computed heteroclinic trajectory $Q(u)$ (solid curves) with the asymptotic approximation in \eqref{eq:twoterm} (dashed curves) for $c=0$ (black/grey), $c=0.1$ (red), $c=0.2$ (green), $c=0.3$ (blue), and $c=0.4$ (pink). (e,f) Comparison of the numerically-computed $c(\kappa)$ relationship (black curve) with its asymptotic approximation in \eqref{eq:ckcsmall} (red dashed curve).
  \label{fig:tw}}
\end{figure}

 As a first example of common reaction-diffusion combinations examined in the context of travelling waves, we consider the \textit{Generalised Porous-Fisher} (GPF) model \cite{murray, aronson1980density, witelski1995merging, mccue2019hole, de1998travelling, de1991travelling, needham1999reaction}, which generalises the common Fisher-Kolmogorov reaction-diffusion model \cite{fisher1937wave, kolmogorov1937etude} to include nonlinear diffusion:
\begin{equation}
D(u)=u^r, \qquad R(u)=u(1-u),\qquad r>-1.\label{eq:GPF}
\end{equation}
From   \eqref{eq:twoterm}, we obtain
\begin{equation}
\xi(u) = \sqrt{\frac{2[1-(r+3)u^{r+2}+(r+2)u^{r+3}]}{(r+2)(r+3)}}. 
\end{equation}
While an explicit solution for $\int_u^1 \xi(s)\mathrm{d}s$ is not possible for general $r$, employing \eqref{eq:Usmall} implies that for $u\to0^+$,
\begin{equation}
\begin{aligned}
z\sim -&\sqrt{\frac{(r+2)(r+3)}{2(r+1)^2}}u^{r+1}\left[1+ \frac{(r+1)(r+3)u^{r+2}}{2(2r+3)} \right.
\\
&\left.+cJ(r)\sqrt{\frac{(r+2)(r+3)}{2}}\left( 1-\frac{(r+1)u}{(r+2)J(r)}+\frac{(r+1)(r+3)u^{r+2}}{2(2r+3)} \right)\right],  \label{eq:GPFz}
\end{aligned}
\end{equation}
where
\begin{equation}
J(r) = \int_0^1 \sqrt{1-(r+3)u^{r+2}+(r+2)u^{r+3}}\,\mathrm{d}u
\end{equation}
and can be easily computed numerically. Additionally, as $u\to1^-$, the GPF model has, from \eqref{eq:gam1}, the far-field approximation
\begin{equation}
u\sim 1-\exp\left[\left(1-\frac{c}{2}\right)\left(z-Z_0\right)\right] ~~\text{ for all } r.
\end{equation}
Using both of these asymptotic approximations, we see in Figure \ref{fig:tw} good agreement with the numerically-determined travelling wave solution. While the two-term approximation of $Q(u)$ via \eqref{eq:twoterm} must be numerically integrated, we also see good agreement between the two-term approximation and the numerically-determined heteroclinic trajectory for small values of $c$.

To contrast with the GPF model, we now consider a new reaction-diffusion model that exhibits travelling waves over a finite compact interval:
\begin{equation}
D(u)=u^{-s}, \qquad R(u)=u^{s}, \quad 0<s<1 \quad\implies \xi(u)=\sqrt{2(1-u)}.\label{eq:FDSR}
\end{equation}
We refer to this reaction-diffusion model as the \textit{Fast Diffusion, Slow Reaction} (FDSR) model, based on the kinetics near $u=0$. While we do not attempt to provide a physical justification for the FDSR, the topics of fast degenerate diffusion are often studied in the context of travelling wave solutions \cite{aronson1980density, murray, mccue2019hole}. The exact solution for $Q(u)$ is shown in \eqref{eq:cst} with $A=1$; however, obtaining $u(z)$ for the FDSR model cannot be explicitly obtained. Using  \eqref{eq:Usmall} and \eqref{eq:gam0}, we have that for $c\ll1$,
\begin{equation}
\begin{aligned}
&z\sim-\frac{u^{1-s}}{\sqrt{2}(1-s)}\left[1+\frac{(1-s)u}{2(2-s)}+\frac{c\sqrt{2}}{3}\left(1-\frac{(1-s)u}{2-s}\right)\right], \quad u\to0^+ ,\\
&z\sim Z_0+\sqrt{2(1-u)}+\frac{2c(1-u)}{3}, \quad u\to1^-, \label{eq:FDSRz}
\end{aligned}
\end{equation}
indicating that the FDSR travelling waves do indeed evolve over a finite interval. As shown in Figure \ref{fig:tw}, these asymptotic approximations agree well with the numerically-determined travelling wave solution. Furthermore, the approximation of $Q(u)$ via \eqref{eq:twoterm} agrees very well with the exact solution shown in \eqref{eq:cst}.

For both the GPF and FDSR models, we can also determine the approximate relationship between $c$ and $\kappa$ when $c\ll1$ via \eqref{eq:ck0}:
\begin{equation}
\text{GPF: } \kappa \sim \frac{c\sqrt{(r+2)(r+3)}}{\sqrt{2}-cJ(r)\sqrt{(r+2)(r+3)}}; \qquad 
\text{FDSR: } \kappa \sim \frac{3c}{3\sqrt{2}-2c}. \label{eq:ckcsmall}
\end{equation}
As seen in Figure \ref{fig:tw}, these approximations for $\kappa(c)$ agree well with the numerically computed values of $\kappa$.

\section{Approximating Reaction-Diffusion-Stefan travelling waves for $\kappa\gg 1$}\label{sec:largeC}

Now knowing the the approximations of $u(z)$ and $Q(u)$ in the limit where $c\ll 1$, i.e. $\kappa\ll1$, a natural extension is to examine travelling waves of the Reaction-Diffusion-Stefan model when the Stefan-like parameter $\kappa$ is large. It is well-reported that behind a critical wavespeed, $c=c^*$, travelling waves of reaction-diffusion models are no longer compactly supported, implying that $\kappa \to \infty$ as $c\to c^{*-}$ \cite{needham1999reaction, de1991travelling, de1998travelling, el2019revisiting, fadai2020new}. In the context of the heteroclinic trajectory $Q(u)$, which solves \eqref{eq:dvdu}, we have that $Q(0)=0$ for $c\ge c^*$. While this critical wavespeed is normally referred to in the literature as the \textit{minimum} wavespeed for which smooth travelling waves exist \cite{murray, sherratt1990models, sherratt1996nonsharp, aronson1980density, witelski1995merging, garduno1994approximation, sanchezgarduno1995traveling, de1991travelling, de1998travelling, needham1999reaction}, we will interpret $c=c^*$ as the \textit{maximum} wavespeed which produce travelling waves with compact support. Additionally, the heteroclinic trajectory $Q(u)$ will ``flatten'' and decrease in absolute magnitude as $c$ increases, due to the key result that travelling waves become more diffuse and spread-out as the wavespeed increases \cite{murray}. We will first show what conditions on $R(u)D(u)$ are required for $c^*=\infty$, as well as the approximations of $Q(u)$ for $c\gg1$. Following this analysis, we will examine instances of $R(u)D(u)$ in which $c^* $ is finite, as well as approximations of $Q(u)$ and $u(z)$ for $c\to c^{*-}$.

To determine necessary conditions for $c^*=\infty$, we first examine a desingularised version of equations \eqref{eq:flux}--\eqref{eq:flux2} via the change of variables
\begin{align}
&\frac{\mathrm{d}u}{\mathrm{d}\zeta}:=D(u)\frac{\mathrm{d}u}{\mathrm{d}z} =Q(\zeta),
\\
&\frac{\mathrm{d}Q}{\mathrm{d}\zeta} =-cQ(\zeta)-R(u)D(u).
\end{align}
This change of variables allows us to examine the equilibrium $(u,Q)=(0,0)$ without the issue of having degenerate $D(u)$ as $u\to0^+$, while leaving $Q(u)$ unchanged \cite{murray, sanchezgarduno1995traveling}. Furthermore, it immediately follows that in order for $(u,Q)=(0,0)$ to be an equilibrium of the desingularised system, we must have $R(0)D(0)=0$. Thus, for $R(0)D(0)\ne0$, no finite value of $c$ produces a heteroclinic trajectory with $Q(0)=0$.

However, having $R(0)D(0)=0$ is a necessary but not sufficient condition for $c^*$ to be finite. By examining the Jacobian $\mathbb{J}$ of the desingularised system at $(u,Q)=(0,0)$, we find that
\begin{equation}
\mathbb{J}_{(0,0)}=\begin{bmatrix}
0 & 1 \\
-\alpha & -c
\end{bmatrix}, ~~\text{where}~~ \alpha=\left.\frac{\mathrm{d}}{\mathrm{d}u}[D(u)R(u)]\right|_{u=0}. \label{eq:jac}
\end{equation}
 However, this local analysis fails when $|\alpha|=\infty$, implying that $Q(0)\ne0$ and providing a second case for when $c^*=\infty$. To summarise, the critical wavespeed $c^*=\infty$ when at least one of the two following conditions hold:
 \begin{equation}
 c^*=\infty \iff R(0)D(0)\ne 0 ~~\text{ or }~~ \left.\frac{\mathrm{d}}{\mathrm{d}u}[D(u)R(u)]\right|_{u\to 0^+}=\pm\infty. \label{eq:CstarInf}
 \end{equation}
 
 \subsection{Approximating the heteroclinic trajectory $Q(u)$ for $c\gg 1$}
 
 We first examine the heteroclinic trajectory $Q(u)$ for $c\gg1$, assuming that the conditions shown in \eqref{eq:CstarInf} hold. By rescaling $Q(u)=c\phi(u)$,   \eqref{eq:dvdu} becomes 
 \begin{equation}
 -c^{-2}\phi(u)\frac{\mathrm{d}\phi(u)}{\mathrm{d}u} = \phi(u)+R(u)D(u), \qquad \phi(1)=0, \qquad \lim_{u\to0^+} \phi(u)=-\frac{c^2}{\kappa}. \label{eq:LargeCode}
 \end{equation}
 Since the left hand side of   \eqref{eq:LargeCode} is $o(1)$, we anticipate that the boundary conditions will not necessarily be satisfied without additional rescaling of variables. Therefore, we first consider the \textit{outer solution} of \eqref{eq:LargeCode} by expanding $\phi(u)=\phi_0(u)+c^{-2}\phi_1(u)+\mathcal{O}(c^{-4})$, where we determine that 
\begin{equation}
\phi_0(u)=-R(u)D(u) ~~\text{ and }~~ \phi_1(u)=-R(u)D(u) \frac{\mathrm{d}}{\mathrm{d}u}[D(u)R(u)];
\end{equation}
hence, the outer solution for $Q(u)$ is 
\begin{equation}
Q_{outer}(u) \sim -\frac{R(u)D(u)}{c}\left\lbrace 1+\frac{1}{c^2} \frac{\mathrm{d}}{\mathrm{d}u}[D(u)R(u)]  \right\rbrace, \qquad c\gg1.
\end{equation}
There are two problems that can arise in the outer solution. The first is if $\mathcal{R} =R(1)D(1)> 0,$  implying that a boundary layer near $u=1$ must exist to satisfy the boundary condition $\phi(1)=0$. By rescaling $u=1-c^{-2}v$ and denoting $\phi = \psi(v)$ in the boundary layer, we obtain the leading-order ODE
 \begin{equation}
\psi(v) \frac{\mathrm{d}\psi(v)}{\mathrm{d}v} = \psi(v)+\mathcal{R} , \qquad \psi(0)=0,\qquad  \lim_{v\to\infty} \psi(v)=-\mathcal{R}, \label{eq:BL}
 \end{equation}
which has the solution $\displaystyle \psi(v) = -\mathcal{R} \left\lbrace1+W\left[-
\exp\left(-1-v\right)\right]\right\rbrace$ and $W(\cdot)$ is the Lambert-W function.

The second problem that can arise is due to the conditions that can hold in \eqref{eq:CstarInf} as $u\to0^+$. As a result, the outer solution no longer remains a well-ordered asymptotic expansion, which is resolved by rescaling   \eqref{eq:LargeCode} as $u\to 0^+$. By assuming $R(u)D(u)\sim Bu^q$ for $u\to 0^+$, where $q\in(-1,1)$, a balance of all terms in \eqref{eq:LargeCode} is achieved via the rescalings
\begin{equation}
u=c^{-\frac{2}{1-q}}\chi, \quad Q=c^{-\frac{1+q}{1-q}}\Phi(\chi),
\end{equation}
which yields the leading-order ODE in the boundary layer near $u=0$:
\begin{equation}
-\Phi(\chi)\frac{\mathrm{d}\Phi(\chi)}{\mathrm{d}\chi} = \Phi(\chi)+B\chi^q, \qquad \Phi(\chi)\sim-B\chi^q ~~\text{as}~~ \chi\to\infty. \label{eq:BL0}
\end{equation}
A special case of   \eqref{eq:BL0}  is when $q=0$, where the solution is $\Phi(\chi)\equiv -B$ and $Q(u)$ therefore does not change in this boundary layer. While   \eqref{eq:BL0} does not have an explicit solution for general $q\in(-1,1)$, we are still able to provide some insight about the composite leading-order approximation of $Q(u)$ for large  $c\gg1$:

\begin{equation}
Q_\text{comp}(u) \sim-\frac{R(u)D(u)-Bu^q+\mathcal{R} \,W\left[-
\exp\left(c^2u-c^2-1\right)\right]  }{c}+c^{-\frac{1+q}{1-q}}\Phi\left(c^{\frac{2}{1-q}}u\right). \label{eq:QComp}
\end{equation}
In particular, we have that 
\begin{equation}
\kappa\sim \frac{c^{\frac{2}{1-q}}}{-\Phi(0)},\qquad c\gg1,
\end{equation}
which provides a leader-order power law relationship between $\kappa$ and $c$ for large wavespeeds.

 \subsection{Approximating the heteroclinic trajectory $Q(u)$ for $c\to c^{*-}$}
 \label{sec:cstar}
As shown in \eqref{eq:CstarInf}, travelling waves in the Reaction-Diffusion-Stefan model can only have a bounded range of wavespeeds, i.e. $c\in[0,c^*]$, if $R(0)D(0)=0$ and $[R(u)D(u)]'$ is finite at $u=0$. From   \eqref{eq:jac}, when $c=c^*, Q(0)=0$ and $\mathbb{J}_{(0,0)}$ has eigenvalues
\begin{equation}
\lambda_{\pm} = \frac{-c^* \pm\sqrt{c^{*2}-4\alpha}}{2}, ~~\text{where}~~  \alpha=\left.\frac{\mathrm{d}}{\mathrm{d}u}[D(u)R(u)]\right|_{u=0}.
\end{equation}
It immediately follows that if $\alpha>0$, $(u,Q)=(0,0)$ changes from a stable spiral point to a saddle point at $c^*=2\sqrt{\alpha}$, thereby providing an explicit expression for the critical wavespeed \cite{murray,sanchezgarduno1995traveling, de1991travelling}. As we impose that $R(u)D(u)\ge0$, we do not consider the case where $\alpha<0$. 

For $\alpha=0$, the eigenvalues of $\mathbb{J}_{(0,0)}$ are $(0,-c^*)$ for all positive wavespeeds, resulting in a fixed point with degenerate stability. Therefore, we cannot use linear stability analysis to determine $c^*$ and instead define the critical wavespeed as the minimum value of $c$ that solves   \eqref{eq:dvdu} and has $Q(0)=0$. Additionally, from the eigenvalues of $\mathbb{J}_{(0,0)}$, we have that $Q'(0)=-c^*$ at the critical wavespeed. While neither $Q(u)$ or $c^*$ can be explicitly determined for all choices of $R(u)D(u)$ with $\alpha$ finite, we are nevertheless interested in examining both $Q(u)$ and $\kappa(c)$ as $c\to c^{*-}$. To do this, we set $c=c^*-\varepsilon$, where $\varepsilon\ll 1$, whereby   \eqref{eq:dvdu} becomes

\begin{equation}
 -Q(u)\frac{\mathrm{d}Q(u)}{\mathrm{d}u} = (c^*-\varepsilon)Q(u)+R(u)D(u), \qquad Q(1)=0. \label{eq:dvdu2}
\end{equation}
By expanding $Q(u)$ as a regular perturbation expansion in $\varepsilon$, i.e. 
$Q(u) = \Psi_0(u) + \varepsilon \Psi_1(u) + \mathcal{O}(\varepsilon^2)$, we obtain, from \eqref{eq:dvdu2}, the following two-term equations:
\begin{align}
&\mathcal{O}_s(1): \quad -\Psi_0(u) \frac{\mathrm{d}\Psi_0(u)}{\mathrm{d}u} =c^*\Psi_0(u)+ R(u)D(u),   \quad &\Psi_0(1)=0; \label{eq:O1e}
\\
&\mathcal{O}_s(\varepsilon): \quad -\frac{\mathrm{d}}{\mathrm{d}u}[\Psi_0(u) \Psi_1(u)] =c^*\Psi_1(u)-\Psi_0(u), \quad &\Psi_1(1)=0. \label{eq:Oce}
\end{align}
To determine how $\Psi_1(u)$ relates to $\Psi_0(u)$, we will assume that both $c^*$ and $\Psi_0(u)$ are known, noting that by its construction, $\Psi_0(0)=0$ and $\Psi_0'(0)=-c^*$. Furthermore, from   \eqref{eq:Oce}, we have that
\begin{equation}
 \frac{\mathrm{d}\Psi_1(u)}{\mathrm{d}u} + \left( \frac{1}{\Psi_0(u)}\frac{\mathrm{d}\Psi_0(u)}{\mathrm{d}u} + \frac{c^*}{\Psi_0(u)}\right) =1;
\end{equation}
thus, by multiplying by the integrating factor
\begin{equation}
\Lambda(u) := -\Psi_0(u)\exp\left(\int^u\frac{c^*}{\Psi_0(s)} \,\mathrm{d} s\right) \ge0, \label{eq:Lambda}
\end{equation}
we obtain
\begin{equation}
\Psi_1(u) = -\frac{1}{\Lambda(u)}\int_u^1\Lambda(s)\mathrm{d} s.
\end{equation}
Noting that $\Lambda(u) \sim c^*$ as $u\to0^+$, we determine that as $c\to c^{*-}$,
\begin{equation}
Q(u) \sim  \Psi_0(u)-\left(\frac{c^*-c}{\Lambda(u)}\right)\int_u^1\Lambda(s)\mathrm{d} s ~~\text{and}~~ \kappa \sim \frac{c^*c}{(c^*-c)\int_0^1\Lambda(s)\mathrm{d} s}. \label{eq:Qcrit}
\end{equation}

\subsection{Comparison between asymptotic approximations and numerical solutions}
\begin{figure}
\begin{center}
\includegraphics[width= .9\textwidth]{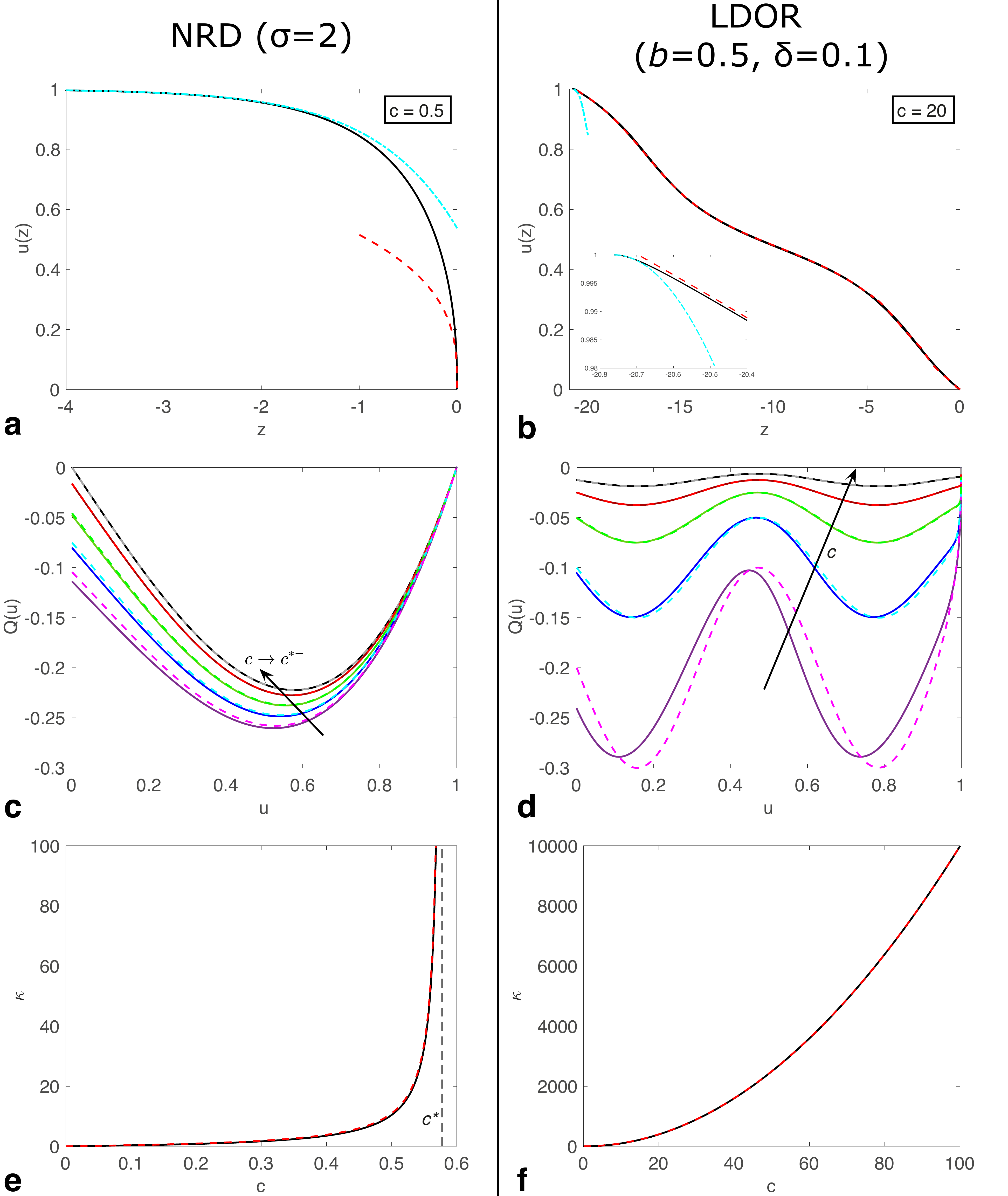}
\end{center}
\vspace{-0.2in} 
 \caption{\textbf{Travelling waves arising in two models of the RDS model and asymptotic approximations valid for $\kappa\gg1$.} Left: NRD model \eqref{eq:NRD} with $\sigma=2$; right: LDOR model \eqref{eq:LDOR} with $b=0./5$ and $\delta=0.1$. (a,b) Numerically-computed travelling waves (black curve) compared with approximations for $u\to0^+$ \eqref{eq:NRDu0},  \eqref{eq:LDORu0} (red dashed curve) and $u\to1^-$ \eqref{eq:NRDu1}, \eqref{eq:LDORu1} (cyan dot-dashed curve), with (a) $c=0.5$ (b) $c=20$. (c,d) Comparisons of the numerically-computed heteroclinic trajectory $Q(u)$ (solid curves) with the asymptotic approximation in \eqref{eq:NRDq},\eqref{eq:LDORq} (dashed curves) for (c) $c=c^*=1/\sqrt{3}$ (black/grey), $c=0.55$ (red), $c=0.5$ (green), $c=0.45$ (blue), and $c=0.4$ (pink); (d) $c=80$ (black/grey), $c=40$ (red), $c=20$ (green), $c=10$ (blue), and $c=5$ (pink). (e,f) Comparison of the numerically-computed $c(\kappa)$ relationship (black curve) with its asymptotic approximation in \eqref{eq:NRDk},\eqref{eq:LDORk} (red dashed curve).
  \label{fig:tw2}}
\end{figure}

With the asymptotic approximations of the heteroclinic orbit $Q(u)$ determined when $\kappa\gg 1$, we now compare numerically-obtained solutions of \eqref{eq:dvdu} for specific choices of reaction kinetics and diffusivities. Another common generalisation of the Fisher-Kolmogorov model is the \textit{Newman Reaction-Diffusion} (NRD) model \cite{newman1983long, witelski1995merging}, in which
\begin{equation}
D(u)=u^\sigma, \qquad R(u)=u(1-u^\sigma),\qquad \sigma>0. \label{eq:NRD}
\end{equation}
The NRD model has an explicitly solvable travelling wave front for a finite critical wavespeed $c^*$, which can be written using the heteroclinic trajectory
\begin{equation}
Q(u)=-\frac{u(1-u^\sigma)}{\sqrt{\sigma+1}}  \qquad c^* = \frac{1}{\sqrt{\sigma+1}}, \qquad u(z) = \left[1-\exp\left(\frac{\sigma z}{\sqrt{\sigma+1}}\right)\right]^{\frac{1}{\sigma}}.
\end{equation}
From   \eqref{eq:Lambda}, we have that
\begin{equation}
\Lambda(u) = \frac{1}{\sqrt{\sigma+1}}(1-u^\sigma)^{\frac{1+\sigma}{\sigma}}, 
\end{equation}
implying, from \eqref{eq:Qcrit}, that  as $c\to c^{*-}$,
\begin{equation}
Q(u) = u^\sigma \frac{\mathrm{d}u}{\mathrm{d}z} \sim -\frac{u(1-u^\sigma)}{\sqrt{\sigma+1}} - \frac{\left(1-c\sqrt{\sigma+1}\right)\left[\mathrm{B}\left(\sigma^{-1},\sigma^{-1}+2\right)-\mathrm{B}_{u^\sigma}\left(\sigma^{-1},\sigma^{-1}+2\right)\right]}{\sigma\sqrt{\sigma+1}\left(1-u^\sigma\right)^{\frac{\sigma+1}{\sigma}}}, \label{eq:NRDq}
\end{equation}
where B$(p,q)$ and B$_x(p,q)$ are the complete and incomplete Beta functions, respectively \cite{chaudhry1997extension}. By evaluating   \eqref{eq:NRD} at $u=0$, we obtain the approximate relationship between $\kappa$ and $c$ as $c\to c^{*-}$:
\begin{equation}
\kappa \sim \frac{c\sigma\sqrt{\sigma+1}}{(1-c\sqrt{\sigma+1})\mathrm{B}\left(\sigma^{-1},\sigma^{-1}+2\right)},\qquad c\to\left(\frac{1}{\sqrt{\sigma+1}}\right)^-. \label{eq:NRDk}
\end{equation}
While  \eqref{eq:NRDq} cannot be explicitly solved for $u(z)$ for all values of $\sigma$, we can expand   \eqref{eq:NRDq} for $u\to 0^+$ to determine the shape of the leading edge of the travelling wave:
\begin{equation}
u(z) \sim \left[(-z)\left(1-c\sqrt{\sigma+1}\right)\left(\frac{\sqrt{\sigma+1}}{\sigma}\right)\mathrm{B}\left(\sigma^{-1},\sigma^{-1}+2\right)\right]^\frac{1}{\sigma+1}, \qquad u\to0^+. \label{eq:NRDu0}
\end{equation}
This expansion confirms that $u \propto (-z)^{1/(\sigma+1)}$ for all $c<c^*$. Furthermore, by expanding  \eqref{eq:NRDq} for $u\to 1^-$, we obtain the far-field behaviour of $u(z)$:
\begin{equation}
u(z) \sim 1-\exp\left[(z-Z_0)\left(\frac{\sigma}{\sqrt{\sigma+1}}\right)\left(1+\frac{1-c\sqrt{\sigma+1}}{1+2\sigma}\right)\right], \qquad u\to 1^-. \label{eq:NRDu1}
\end{equation} 
In Figure \ref{fig:tw2}, we see that these leading-order approximations for $Q(u)$ and $\kappa(c)$ are in good agreement with the numerical solution of the NRD model for $c\to c^{*-}$. While approximations of $u(z)$ are less accurate near $u=0$, we note that \eqref{eq:NRDu0} is only a leading-order approximation and further agreement can be obtained with the incorporation of higher-order terms in \eqref{eq:NRDq}. Nevertheless, the far-field approximation as $u\to1^-$ agrees well with the numerical solution of the NRD model.

Our final reaction-diffusion model we will consider is the \textit{Linear Diffusion, Oscillatory Reaction} (LDOR) model, in which
\begin{equation}
D(u)=1, \qquad R(u)=1+b\sin\left(\frac{u}{\delta}\right),\qquad  b<1, \quad \delta>0. \label{eq:LDOR}
\end{equation}
Like the FDSR model in Section \ref{sec:smallc}, we do not attempt to provide a physical justification for the LDOR model, but instead note that, since $R(0)D(0)=1$, the critical wavespeed of the LDOR model is $c^*=\infty$. We can therefore use \eqref{eq:QComp} to determine the approximate heteroclinic trajectory when $c\gg 1$:
\begin{equation}
Q(u)\sim -\frac{1 }{c}\left\lbrace 1+b\sin\left(\frac{u}{\delta}\right)+\mathcal{R} \,W\left[-
\exp\left(c^2u-c^2-1\right)\right]\right\rbrace, \qquad \mathcal{R}  = 1+b \sin\left(\frac{1}{\delta}\right). \label{eq:LDORq}
\end{equation}
Additionally, we determine that for $c^2(1-u)\gg1$, i.e. for $u$ far away from 1, the travelling wave has the approximation
\begin{equation}
\begin{aligned}
z\sim &-c\int_0^u \frac{\mathrm{d} s}{R(s)}
\\
&=-\frac{2\delta c}{\sqrt{1-b^2}}\left\lbrace\tan^{-1}\left[\frac{\tan\left(\frac{u}{2\delta}\right)+b}{\sqrt{1-b^2}}
\right]-\sin^{-1}(b)+k\pi\right\rbrace, \qquad k\in\mathbb{Z},
\end{aligned}\label{eq:LDORu0}
\end{equation}
where $k\pi$ is chosen so that $u(z)$ is continuous on $[0,1]$. For $c^2(1-u)\ll 1$, $u(z)$ is described predominantly by the boundary layer of the heteroclinic orbit, implying that
\begin{equation}
z\sim Z_0-\int_0^u \frac{D(1)\mathrm{d} s}{\mathcal{R}\sqrt{2(1-u)}}
= Z_0-\frac{\sqrt{2(1-u)}}{\mathcal{R}}.
\label{eq:LDORu1}
\end{equation}
Finally, since $Q(0)\sim -c^{-1}$, we determine that
\begin{equation}
\kappa \sim c^2, \qquad c\gg1.\label{eq:LDORk}
\end{equation}
As seen in Figure \ref{fig:tw2}, these leading-order approximations for $u(z), Q(u)$ and $\kappa(c)$ are in good agreement with the numerical solution of the LDOR model for $c\gg1$. Specifically, we note that \eqref{eq:LDORu0} is valid for the majority of the $u(z)$ travelling, since the boundary layer near $u=1$ is $\mathcal{O}_s(c^{-2})$ in height (see inset of Figure \ref{fig:tw2}b). Therefore, we conclude that our leading-order asymptotic approximations for large-$\kappa$ travelling waves in the Reaction-Diffusion-Stefan model framework are valid for both $c^*$ finite and infinite.

\section{Conclusions}

In this work, we consider compactly supported travelling waves arising from a general reaction-diffusion model coupled with a Stefan-like boundary condition at the leading edge of the front. This Reaction-Diffusion-Stefan (RDS) model employs a nonlinear diffusivity, $D(u)$, as well as a nonlinear, non-negative reaction term, $R(u)$. Using travelling wave coordinates, we transform the model into a single nonlinear differential equation, whose solution is the heteroclinic trajectory of the travelling wave in the phase plane. Unlike other reaction-diffusion models, compactly support travelling waves arise in the RDS model for a \textit{range} of wavespeeds, as opposed to a single, critical wavespeed. In the limiting regime where the wavespeed is small, i.e. $c\ll1$, we obtain a good approximation of the heteroclinic trajectory of the travelling wave, which can also be used to relate the speed of the front to the Stefan-like parameter, $\kappa$. We also determine the approximate form of the travelling wave near its leading edge ($u\to0^+$), as well as when $u\to1^-$. The key result of this analysis is that in order for travelling waves in the RDS model to have compact support, both $R(u)D(u)$ and $D(u)$ must be $o(u^{-1})$ for $u\to0^+$. Furthermore, travelling waves that evolve over a finite compact interval require that $R(u)=\mathcal{O}((1-u)^\gamma)$ as $u\to1^-$, where $\gamma\in[0,1)$. Finally, travelling waves in the RDS model evolving on a semi-infinite domain must have $R(u)=\mathcal{O}_s(1-u)$ as $u\to1^-$. 

We also perform a similar asymptotic analysis of travelling wave solutions of the RDS model for wavespeeds approaching a critical wavespeed,  $c^*$. This threshold wavespeed provides a bound for when travelling waves fail to provide compact support. Based on the behaviour of $R(u)D(u)$ as $u\to0^+$, we provide the necessary conditions for which $c^*$ is finite. For both $c^*=\infty$ and $c^*$ finite, we obtain approximations of the heteroclinic trajectory and the relationship between $c$ and $\kappa$ as $c\to c^{*-}$. To validate these asymptotic approximations, we examine various choices of $R(u)$ and $D(u)$ in the RDS model that are commonly employed in the literature. In all cases, we find that our asymptotic approximations for the heteroclinic trajectory, the relationship between wavespeed and Stefan-like parameter, and the travelling wave itself all agree well with their corresponding numerical solutions.

Further extensions of the RDS model can also be considered. For instance, one could relax the assumption that $R(u)\ge0$ for $u\in[0,1]$, as is done with the Allee-type reaction kinetics \cite{li2019travelling, johnston2017co}. For this class of reaction kinetics, the asymptotic analysis described in this work is no longer valid, as the heteroclinic trajectory in the phase plane is no longer strictly negative. While other methods have been proposed to mitigate these issues in related reaction-diffusion models \cite{li2019travelling}, it is unclear how these techniques can be used in the RDS modelling framework. In addition to negative reaction kinetics, other extensions can be incorporated into the RDS modelling framework, including nonlocal reactions \cite{billingham2020slow} and multiple-species reaction-diffusion equations \cite{mimura2000reaction, el2020sharp}. We leave these extensions for future consideration. 
\newpage

\bibliographystyle{ieeetr}
\bibliography{RD}
 
\end{document}